# Room-temperature antiferromagnetic memory resistor

X. Marti[1,2,3,*], I. Fina[4,5], C. Frontera[4], Jian Liu[6], P. Wadley[3,12], Q. He[7], R.J. Paull[1], J.D. Clarkson[1], J. Kudrnovský[8], I. Turek[2,9], J. Kuneš[3], D. Yi[6], J.-H. Chu[6], C.T. Nelson[10], L. You[11], E. Arenholz[7], S. Salahuddin[11], J. Fontcuberta[4], T. Jungwirth[3,12], R. Ramesh[1,6,10]

[1] Department of Materials Science and Engineering, University of California, Berkeley, California 94720, USA

[2] Department of Condensed Matter Physics, Faculty of Mathematics and Physics, Charles University, 12116 Praha 2, Czech Republic

[3] Institute of Physics ASCR, v.v.i., Cukrovarnická 10, 162 53 Praha 6, Czech Republic

[4] Institut de Ciència de Materials de Barcelona, ICMAB-CSIC, Campus UAB, E-08193 Bellaterra, Spain

[5] Max Planck Institute of Microstructure Physics, Weinberg 2, D-06120 Halle, Germany

[6] Materials Science Division, Lawrence Berkeley National Laboratory, Berkeley, California 94720, USA

[7] Advanced Light Source, Lawrence Berkeley National Laboratory, Berkeley, California 94720, USA

[8] Institute of Physics ASCR, v.v.i., Na Slovance 2, 182 21 Praha 8, Czech Republic

[9] Institute of Physics of Materials ASCR, v.v.i., Zizkova 22, 616 62 Brno, Czech Republic

[10] National Center for Electron Microscopy, Lawrence Berkeley National Laboratory, Berkeley, California 94720, USA

[11] Department of Electrical Engineering and Computer Sciences, University of California, Berkeley, Berkeley, California 94720, USA

[12] School of Physics and Astronomy, University of Nottingham, Nottingham NG7 2RD, United Kingdom

* xavi.marti@igsresearch.com

**The bistability of ordered spin states in ferromagnets (FMs) provides the magnetic memory functionality. Traditionally, the macroscopic moment of ordered spins in FMs is utilized to write information on magnetic media by a weak external magnetic field, and the FM stray field is used for reading. However, the latest generation[1] of magnetic random access memories[2] demonstrates a new efficient approach in which magnetic fields are replaced by electrical means for reading and writing. This concept may eventually leave the sensitivity of FMs to magnetic fields as a mere weakness for retention and the FM stray fields as a mere obstacle for high-density memory integration. In this paper we report a room-temperature bistable antiferromagnetic (AFM) memory which produces negligible stray fields and is inert in strong magnetic fields. We use a resistor made of an FeRh AFM whose transition to a FM order 100 degrees above room-temperature[3,4,5] allows us to magnetically set different collective directions of Fe moments. Upon cooling to room-temperature, the AFM order sets in with the direction the AFM moments pre-determined by the field and moment direction in the high temperature FM state. For electrical reading, we use an antiferromagnetic analogue of the anisotropic magnetoresistance (AMR). We report microscopic theory modeling which confirms that this archetypical spintronic effect discovered more than 150 years ago in FMs[6,7] can be equally present in AFMs. Our work demonstrates the feasibility to realize room-temperature spintronic memories with AFMs which greatly expands the magnetic materials base for these devices and offers properties which are unparalleled in FMs.**

In FMs the bistability, i.e., the energy barrier separating two stable directions of ordered spins is partly due to the macroscopic moment and the corresponding dipolar shape anisotropy fields. Another contribution, which in magnetically hard FMs dominates, is from the spin-orbit magnetocrystalline anisotropy energies. Since the relativistic spin-orbit coupling makes no difference between FMs and AFMs and magnetocrystalline anisotropies are even in the microscopic moment direction, the bistability can be readily present even if the collective spin-order is AFM.[8,9]

The magneto-transport counterparts of the relativistic energy anisotropies are the AMR effects. Their presence in AFMs was predicted[10] and experimentally confirmed[11,12,13,14] in tunnel junction devices with an AFM IrMn electrode and a non-magnetic counter-electrode on the other side of the tunnel barrier. The spin-axis of AFM aligned moments can point in different directions and can be detected electrically by the AMR as in FMs.

In ferromagnetic materials, due to their net magnetization, the electron spins can be oriented by a magnetic field larger than the coercive field, whose strength is limited by the magnetic anisotropy fields and rarely exceeds ~1 T. Compensated AFMs, in contrast, have zero net magnetization and thus, unlike FMs, produce no stray magnetic fields. The zero magnetization also implies that, AFMs are extremely weakly sensitive to external magnetic fields. The sensitivity scale of AFMs is set by the ~100 Tesla exchange fields.[15]

Recently the concept of an AFM memory was explored on an atomic-scale.[16] Bistable AFM configurations of a few Fe atom chain were detected using low-temperature, atomically-resolved measurement by a spin-polarized tip of a scanning-tunneling-microscope. Here we demonstrate an AFM-AMR memory which has a simple Ohmic resistor geometry and the magnetic state is stored and detected up to high temperatures. We show that the AFM memory is inert in magnetic fields and produces negligible magnetic stray fields.

Our main results are summarized in Fig. 1. We have grown a 100 nm thick, single-crystalline film of the FeRh AFM on a cubic MgO substrate. The AFM-FM transition occurs in FeRh close to 400 K; at this temperature, we have applied a magnetic field to align its magnetization and the corresponding magnetic moments of FM FeRh along the applied magnetic field. The sample is then field-cooled below room-temperature (200 K) and the magnetic field is removed. In this AFM state with no applied magnetic field we perform a series of 4-probe resistance measurements. The same protocol is repeated several times with the magnetic field applied during field-cooling either along the [100] or [010] substrate crystal direction (see Fig. 1a). The resulting resistances in the AFM state are fully reproducible and the two cooling-field directions define two distinct resistance states of the AFM. They remain distinct not only upon removing the magnetic field but also when warming the AFM up to room-temperature, as shown in Figs. 1b,c.

In Figs. 1d,e we demonstrate that the two AFM memory states are robust against strong magnetic field perturbations. After preparing one of the states by the above cooling-in-field procedure we rotate the sample at room-temperature in a 1 T field and observe a negligible effect on the resistance in either of the two AFM memory states (Fig. 1e). Similar to the rotation experiment, the states are not disturbed by sweeping the magnitude of the magnetic field at a fixed angle (Fig. 1d). In the detailed discussion below we show that the retention in our AFM memory is not disturbed up to the highest fields (9 T) available in our set-up. However, before resuming the detailed experimental analysis, we focus in the following paragraphs on the microscopic physics behind the observed distinct resistance states in our FeRh AFM. In this theoretical section we first recall the fundamentals of the AMR relevant to our experiments and then discuss our quantitative modeling of the effect based on a relativistic density-functional transport theory.[17,18,19,20,21]

As already mentioned in the introduction, conceptually the AMR phenomena are equally present in AFMs as in FMs. Since AMR is an even function of the microscopic magnetic moment vector it is the direction of the spin-axis rather than the direction of the macroscopic magnetization which primarily determines the effect. In collinear FMs the two directions are equivalent. For the staggered spin configuration of compensated AFMs only the spin-axis can be defined while the macroscopic magnetization is zero.

To date, studies of AMR phenomena in AFMs have focused on tunneling devices.[10,11,12,13,14] In analogy to the tunneling AMR in FMs,[22,23,24,25,26,27] the origin of the phenomenon is ascribed to the changes in the equilibrium relativistic electronic structure (density of states) induced by rotating the spin-axis with respect to crystal axes.[10] A recently reported AMR in an Ohmic device fabricated from an AFM $Sr_2IrO_4$ film shares the anisotropic electronic structure origin of the tunneling AMR.[28] In the $Sr_2IrO_4$ AMR experiment, the current was driven through the film in the direction perpendicular to the sample plane while the AFM spin-axis was rotated in the plane. This experimental geometry in which the angle between the spin-axis and current is fixed and the resistance depends on the angle between spin-axis and crystal-axes is referred to as the "crystalline" Ohmic AMR.[7,29] For both the tunneling AMR or the crystalline Ohmic AMR, the quantitative relativistic transport theory would require to combine the calculated density of states anisotropy with the tunneling or scattering matrix elements, respectively. Since a proper modeling of these matrix elements for realistic sample parameters is in general a difficult problem, the theories have focused primarily on assessing the qualitative origin of these AMR phenomena based on the density of states anisotropy calculations.

In our Ohmic FeRh resistor the different directions of the AFM spin-axis are set by a field-cooling procedure in which the magnetic fields are applied either along the [100] or [010] crystal axis of our MgO/FeRh sample. Due to the in-plane cubic symmetry of the structure, the equilibrium density of states of the FeRh film is identical for the AFM spin-axis set by the cooling-field applied along equivalent cubic axes. Unlike the above tunneling or crystalline Ohmic AMR, the equilibrium density of states anisotropy is not the origin of the resistance anisotropy we observe in our FeRh resistor. Instead, the effect originates from the change of the angle of the spin-axis with respect to the electrical current direction. This is akin to the common "non-crystalline" AMR in FMs[7,29] which is the only component contributing to the AMR in polycrystalline samples or when the spin-axis alternates among equivalent symmetry axes of the crystal. The dependence on the relative angle of the spin-axis and current is the phenomenon that Kelvin detected in his more than 150 years old seminal work on the AMR in Fe and Ni.[6]

Since the equilibrium density of states anisotropy of a clean crystal does not provide even a qualitative proxy to the non-crystalline AMR, its modeling remains a challenge despite the long history of the effect and the simple experimental geometry used for its detection. In the leading order, the non-crystalline AMR reflects the dependence of the transport scattering matrix elements of electrons with momentum parallel to current on the angle between the spin-axis and the momentum. The spin-dependence of the momentum scattering is due to the relativistic spin-orbit coupling. A quantitative modeling requires to identify the dominant scattering mechanism and to use a full quantum transport theory framework. In FMs, such a quantitative description has been successfully applied to random metal alloys in which scattering is dominated by the alloy disorder and the relativistic band structure is reliably obtained using the density-functional theory.[17,18,19,20,21] Here we apply analogous formalism to calculate the non-crystalline Ohmic AMR in our FeRh AFM.

Results of our microscopic calculations are presented in Fig. 1f. We employed the relativistic, tight-binding linear muffin-tin orbital (TB-LMTO) density functional theory of the band structure, the Kubo formula description of transport, and the alloy disorder is accounted for within the coherent potential approximation (CPA).[20,21] The AMR plotted in Fig. 1f is defined as a relative difference between the resistivity for the spin-axis parallel and perpendicular to current, AMR=$(R_{s\|j}-R_{s\perp j})/R_{s\|j}$. This applies to both the AFM ground-state of FeRh and an hypothetical zero-temperature FM state which we also consider in our calculations for comparison. Following the chemical composition measurements shown below in Fig. 2, we assume in the calculations a Rh-rich $(Fe_{1-x}Rh_x)Rh$ random alloy and calculate the AMR as a function of x. The experimentally determined non-stoichiometry in our samples corresponds to x ≈ 0.02 - 0.03 (Fe/Rh ~ 0.94 - 0.96). For these values, the calculated AMR ≈ 0.1-1% in the AFM state is in good agreement with the measured difference between the two resistance states. The somewhat larger theoretical than experimental values are expected since the calculations are performed at zero temperature. To confirm the robustness of the theoretically predicted AMR in AFMs we also performed control calculations for FeRh in which disorder and temperature-dependent scattering (on phonons or magnons) is effectively modeled by a finite quasiparticle broadening of the Fermi level states. For residual resistivities similar to the experimental values we consistently obtain AFM-AMR values close to a half of a per cent. The comparison of our calculated FeRh AMRs in the AFM ground-state and the hypothetical zero-temperature FM state highlight the expected equal presence and comparable magnitude of the AMR phenomena in these two distinct ordered spin states. Note that the calculated AFM-AMR even exceeds the FM-AMR at stronger alloy disorder (larger x), as seen on Fig. 1f.

An important evidence is drawn from the calculated sign of the AFM-AMR in our FeRh sample. Similar to the calculated sign of the AMR in the FM state of FeRh and to the AMR sign in common transition metal FMs,[7,17,18,19,20,21] our theory predicts a lower resistivity state for the AFM spin-axis aligned perpendicular to the current. This confirms the scenario, anticipated in Fig. 1a, by which the field cooling procedure sets in the spin-axis direction in the FeRh AFM state. In agreement with the theoretical sign of the AFM-AMR, the ferromagnetically aligned moments along the applied magnetic field at the high temperature rotate clockwise in one spin-sublattice and counter-clock-wise in the other sublattice of FeRh until they reach the AFM configuration at the low temperature with the spin-axis oriented perpendicular to the cooling-field direction. Having established the microscopic physics description of key functionalities of our AFM memory resistor we resume in the remaining paragraphs the more detailed description of our experimental findings.

The high epitaxial quality of our FeRh films is illustrated in Fig. 2. The cross section of the FeRh thin film at the interface with the MgO substrate (Fig. 2a), viewed down the (112) and (100) zone axis of the two respective materials, shows that the FeRh epilayer is fully coherent with the MgO substrate. The micrograph demonstrates nearly perfect ordering of the atoms in the FeRh unit cell. X-ray diffraction patterns (Fig. 2b) display only the {00l} family of peaks of both FeRh and MgO confirming that the samples are textured with no spurious phases. In Fig. 2c we show the wavelength dispersive X-ray spectroscopy analysis of the chemical composition of our FeRh film. We observe a small Fe deficit from the target's nominal 1:1 stoichiometry with Fe/Rh≈0.98-0.96.

Fig. 3 presents magnetic characterization of our FeRh film by the superconducting quantum interference device (SQUID) and vibrating sample magnetometer (VSM) which provides further insight into the behavior our AFM memory. In Fig. 3a we show temperature-dependent magnetization data recorded at different magnetic fields. When warming up, the magnetization displays a clear signature of the transition from the AFM to the FM phase of FeRh. The thermal hysteresis reflects the first order nature of the transition[4]. When increasing the magnetic field, the FM phase is stabilized and the transition occurs at lower temperatures. In the present context, of relevance is that at 400 K, a field of 9 T is strong enough to fully stabilize the FM state. At lower temperatures the system gradually turns into the AFM state.

Complementary magnetic-field-dependent magnetization traces are plotted in Figs. 3b,c. At 400 K, the data are consistent with the FM alignment of Fe spins and identical signal is detected while sweeping the magnetic field along the [100] or [010] crystal directions (Fig. 3c). This confirms the expected in-plane cubic symmetry of our FeRh film deposited on the cubic MgO substrate. We note that the loops were obtained by sweeping the magnetic field between ±9 T at 400 K. Magnetization data measured at room-temperature and 200 K (Fig. 3b) confirm the AFM order. Only a weak residual moment remains in the system as seen from the minor hysteretic contribution to the signal at low (~100 mT) fields. Its origin can be associated with the small departure of our FeRh film from the nominal 1:1 stoichiometry of the fully compensated FeRh AFM (see Fig. 2b) or with the presence of residual ferromagnetic regions in the AFM matrix as is common in first-order transitions. We ascribe the weak variation of the resistance seen in Fig. 1e in the field-rotation measurement to this residual moment. We point out, however, that the residual moment which is reversed at the low, ~100 mT field (Fig. 3b) cannot be responsible for the AMR signal distinguishing the two memory states since these distinct resistance states remain stable up to the largest field (9 T) applied in our experiments. The residual moment has also no marked effect on our AFM memory states as the difference between the corresponding resistances is much larger than the weak variations of the individual angular dependencies of each of the two memory states.

Magnetization data in Figs. 3a-c are further complemented by the temperature-dependent magneto-transport measurements shown in Fig. 3d. At 340 K, the moments are aligned ferromagnetically in the large applied magnetic field and, as a result, we observe a negative and hysteretic magnetoresistance trace in the field-sweep measurement. Similar data are obtained at 320 K. However, the hysteretic field-dependent part of the resistance trace is shifted to higher magnetic fields as we move further away in temperature from the AFM-FM transition. We note that these measurements are consistent with a recent study of similar FeRh/MgO samples which focused on magnetotransport effects near the AFM-FM transition.[4] The aim of our paper is, however, to demonstrate the memory functionality and AMR in the AFM state below the transition. In particular, at room-temperature and at 200 K we do not see any field-dependence of the resistance on the x-y-scale of Fig. 3d, confirming that the FeRh film behaves basically as a compensated AFM with a negligible residual moment at these low temperatures and up to the highest applied magnetic fields.

In Fig. 4 we inspect the 200 K and room-temperature resistance traces in the large field-range and on a fine resistance scale, for the two memory states set up by the cooling-in-field procedures described in Fig. 1. At

200 K we observe a parabolic positive magnetoresistance which reaches ∼1% at 9 T (Fig. 3a). We ascribe it to the ordinary, Lorentz force magnetoresistance which is independent of spin. Indeed, the two traces have just a constant off-set, i.e., cycling our AFM memory between ±9 T at 200 K has no effect on the stability of the two distinct AFM states. This is further confirmed in Fig. 4b where instead of sweeping the magnetic field we rotate the sample in a 9 T applied field. We observe small variations of the resistance which we above ascribed to the residual uncompensated moments. However, as seen in Fig. 4b, the two distinct AFM memory states remain clearly separated even in the 9 T rotating field.

In Fig. 4c we plot room-temperature magnetoresistance traces on a similar fine scale and we observe a different behavior in the strong magnetic fields than at 200 K. The ordinary spin-independent positive magnetoresistance is not visible because of a stronger hysteretic negative magnetoresistance. Comparing to the traces at 320 and 340 K plotted in Fig. 3d it is clear that this, still relatively tiny hysteretic magnetoresistance seen at room-temperature reflects that with increasing temperature and large field the system approaches the transition to the FM coupling of spins. However, at room-temperature and up to the highest applied magnetic field of 9 T the system remains far from the complete transition to the FM order. As further highlighted in Fig. 4d, this implies that we can cycle our AFM FeRh memory between ±9 T fields at room-temperature without disturbing either of the two distinct memory states.

To conclude, we have demonstrated a room-temperature AFM memory and microscopically explained the AFM-AMR effect we employed for a simple Ohmic resistance read-out. Our magnetic memory shows the unique features of AFMs, in particular, it is insensitive to strong magnetic field perturbations and produces negligible stray fields. The magnitude of the AFM-AMR effect in our FeRh sample is sufficient to reproducibly detect the two distinct memory states at room-temperature.

Our work represents the proof of principle demonstration of an AFM memory resistor, yet there is a large room for improving all the functionalities of such a device. Here it is important to note that the magnetic memory concept we introduce is generic to a broad class of spin-orbit coupled AFMs. Our microscopic calculations indicate a route for increasing the magnitude of the AFM-AMR in FeRh and the same theory framework, or direct measurements, can systematically map Ohmic AMRs in a variety of other metallic AFMs. To improve our cooling-in-field procedure for writing, future studies of AFM memories can benefit from the extensive research and development of techniques in FMs for heat-assisted magnetic recording.[2] In our FeRh samples, the temperature and field-strength can still be optimized to achieve the most efficient writing in this material. Other systems which show the high-temperature AFM-FM transition can be considered or, as recently demonstrated in a tunneling IrMn device,[14] heating close to the AFM to paramagnet transition in materials without the AFM-FM transition point may by a viable alternative. A highly attractive approach for AFMs is to use spin-orbit coupling for both reading and writing the state by electrical means. This is a concept which in FMs has recently drawn much attention from both fundamental physics and memory application perspectives[30,31] but in AFMs is yet to be explored. Beyond the electrical means, spins in magnetic materials can be reoriented optically which in AFMs in particular may allow for ultrafast writing schemes.[32]

**Methods**

Thin films were prepared by DC sputtering. MgO(001) substrates were heated up to 550 K in a base pressure of $10^{-8}$ Torr. Subsequently, Ar gas was introduced (3 mTorr) and films were grown using 50 W power at a rate of 1 nm/min using an stoichiometric FeRh target.

X-ray diffraction experiments were carried out using a Panalytical X'Pert Pro diffractometer by Philips. The composition and thickness of the thin films were obtained by variable voltage electron probe microanalysis, using a CAMECA SX-50 electron microprobe equipped with four wavelength-dispersive x-ray spectrometers. X-ray intensities were measured at 10, 12, 15 and 20 keV electron incident energy and they were analyzed with the help of the program STRATAGEM (SAMX, France), which calculates the thickness and composition of a multilayer target by least squares fitting of an analytical x-ray emission model to the experimental data.

Cross sectional sample for TEM was prepared by mechanical polishing followed by Ar ion milling. HAADF STEM imaging was performed on a FEI titan, the TEAM0.5 at the national center for electron. The image shown in fig. 2a is an averaged region from a Scanning Transmission Electron Microscopy (STEM) image taken with a High Angle Annular Dark Field (HAADF) detector. This mode highlights the atom nucleus with intensity corresponding to the atomic number, here Rh(Z=45) appears brightest followed by Fe (Z=26), and Mg (Z=12) in the substrate. The image was averaged laterally over approximately 30 adjacent unit cells.

Magnetic characterization with magnetic fields up to 5 T was carried out in SQUID by Quantum Design using the Reciprocating Sample Option (RSO). Measurements up to 9 T were performed in a vibrating sample magnetometer by Quantum Design.

Transport experiments were 4-probe resistance measurements carried out using the Physical Property Measurement System by Quantum Design in resistivity mode and equipped with a horizontal rotator module. For parallel current (bi-polar constant current ranging from 1 to 10 μA) and magnetic field position θ is defined as 0, along the substrate [100] direction shown.

Kubo formula CPA-TB-LMTO calculations were performed for the cubic, CsCl crystal structure of FeRh. The structure of the AFM state was described as a stacking of four interpenetrating fcc-sublattices along the [111]-direction occupied respectively by Fe[up], Rh, Fe[down], and Rh atoms. Substitutional $Rh_{Fe}$ disorder was placed on both Fe-sublattices in the considered Rh-rich $(Fe_{1-x}Rh_x)Rh$ random alloy. The model of quasiparticle broadening was treated by adding a constant, spin-independent imaginary part of the self-energy (of order $10^{-3}$ - $10^{-2}$ Ry). A large number (~$10^7$) of points in the full fcc Brillouin zone was used to obtain converged AMR values.

## Author contributions

*Sample preparation, R.J.P, J.C., L.Y; scanning transmission electron microscopy, C.T.N.; experiments and data analysis: I.F., C.F., J.H.C., D.Y., J.L., E.A., Q.H; theory, J.K., I.T.; writing and project planning, X.M., T.J., J.F., S.S., R.R.*

## Aknowledgements

*The authors acknowledge the support from the NSF (Nanosystems Engineering Research Center for Translational Applications of Nanoscale Multiferroic Systems, Cooperative Agreement Award EEC-1160504) and DOE.TEM characterization was performed at NCEM, which is supported by the Office of Science, Office of Basic Energy Sciences of the U.S. Department of Energy under Contract No. DE-AC02—05CH11231. J.F. acknowledges financial support from the Spanish Government (Projects MAT2011-29269-C03, CSD2007-00041), Generalitat de Catalunya (2009 SGR 00376); I.F. acknowledges Beatriu de Pinós postdoctoral scholarship (2011 BP-A 00220), the Catalan Agency for Management of University and Research Grants (AGAUR- Generalitat de Catalunya). X.M. acknowledges the Grant Agency of the Czech Republic No. P204/11/P339. T.J. acknowledges support from the ERC Advanced Grant 268066, and Praemium Academiae of the Academy of Sciences of the Czech Republic.*

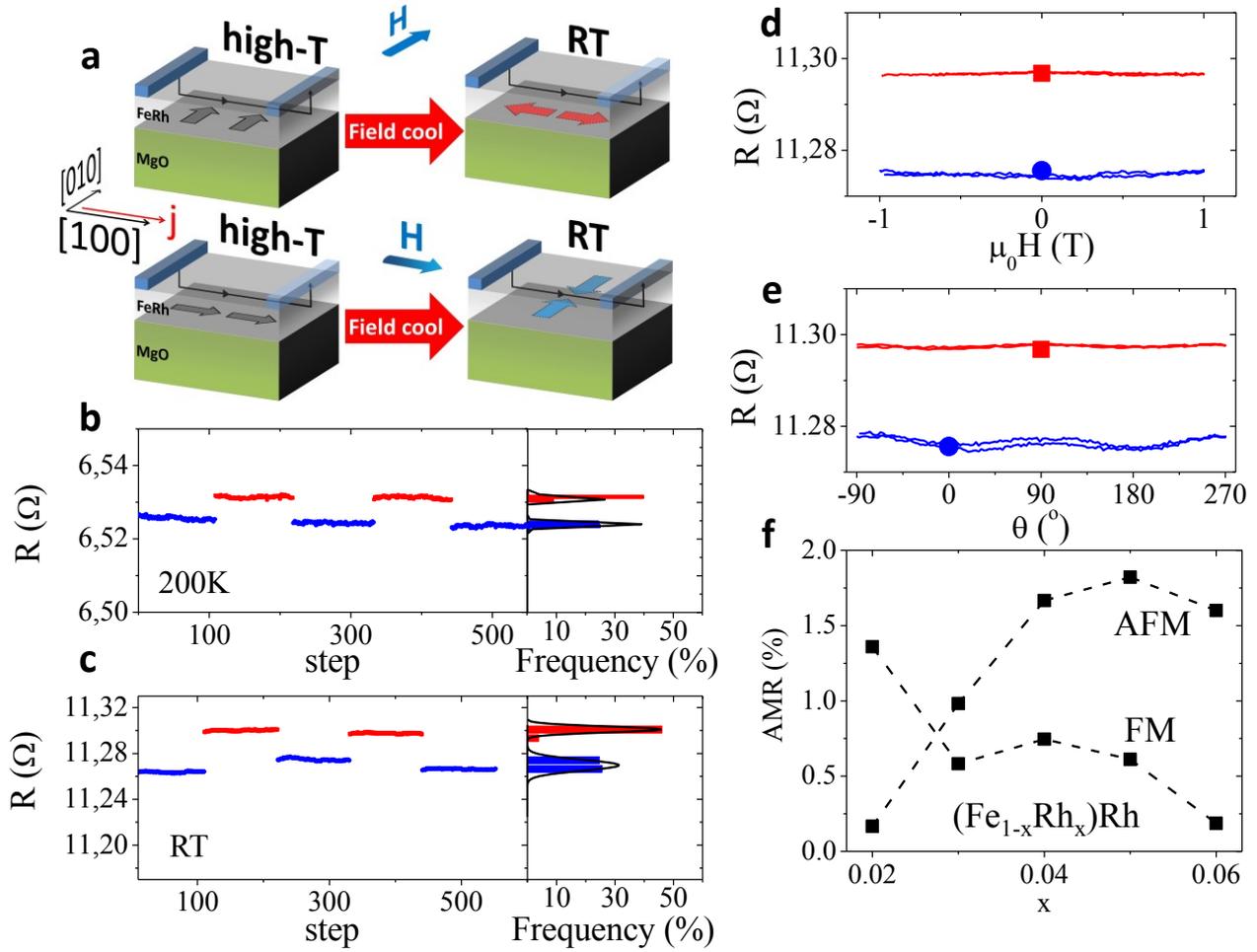

**Figure 1. AFM-AMR memory functionality in an FeRh resistor. a,** Schematic illustration of the AFM FeRh/MgO structure and of the memory writing and reading set up. For writing the sample is cooled in a field ***H*** from a temperature above the AFM/FM transition in FeRh to below the transition temperature (200 K). The resulting AFM spin-axis in the low-temperature memory state depends on the direction of *H* which is either along the [100] or [010] crystal axis. For reading, electrical current *j* is driven between electrical contacts (blue bars) along the [100] direction and the resistance is detected. **b,** Resistance measured at 200 K and zero magnetic field after field cooling the sample with the magnetic field parallel (blue) and perpendicular (red) to the current direction. The two resistance states are clearly distinct. This is highlighted in right panel histograms showing the relative frequency of the measured resistance values. **c,** Same as **b**, at room temperature. **d,** Stability of the two memory states at room temperature tested by measuring the resistance while sweeping a magnetic field between ±1 T applied along the [100] direction. **e,** Same as d, while rotating a 1 T magnetic field. **f,** AMR values calculated for the Rh-rich $(Fe_{1-x}Rh_x)Rh$ random alloy using the Kubo formula CPA-TB-LMTO formalism. The AMR is defined as a relative difference between the resistivity for the spin-axis parallel and perpendicular to current, $AMR=(R_{s\|j}-R_{s\perp j})/R_{s\|I}$. Results are shown for the AFM ground-state and for a hypothetical zero-temperature FM state of FeRh.

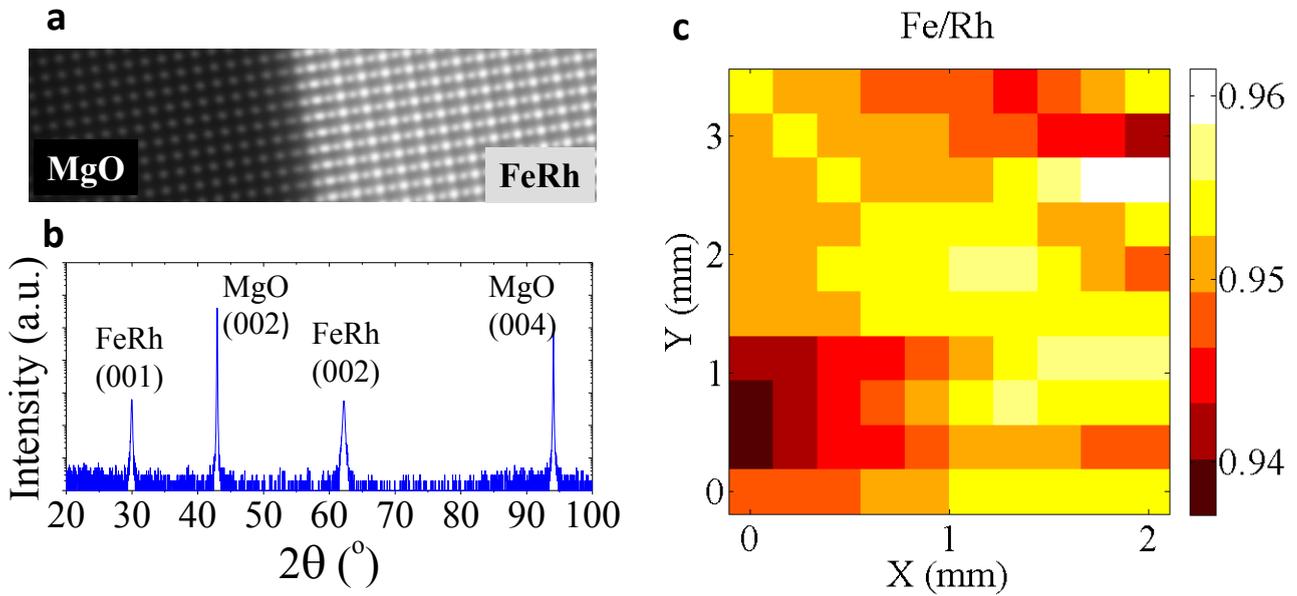

**Figure 2. Structural and compositional characterization of the FeRh film. a,** Cross section image of the FeRh thin film at the interface with the MgO substrate, viewed down the (112) and (100) zone axis of the two respective materials. **b,** θ-2θ scan showing textured films, indicated are the (00l) reflections of the substrate and the film. **c,** Compositional map of the Fe/Rh content of the film.

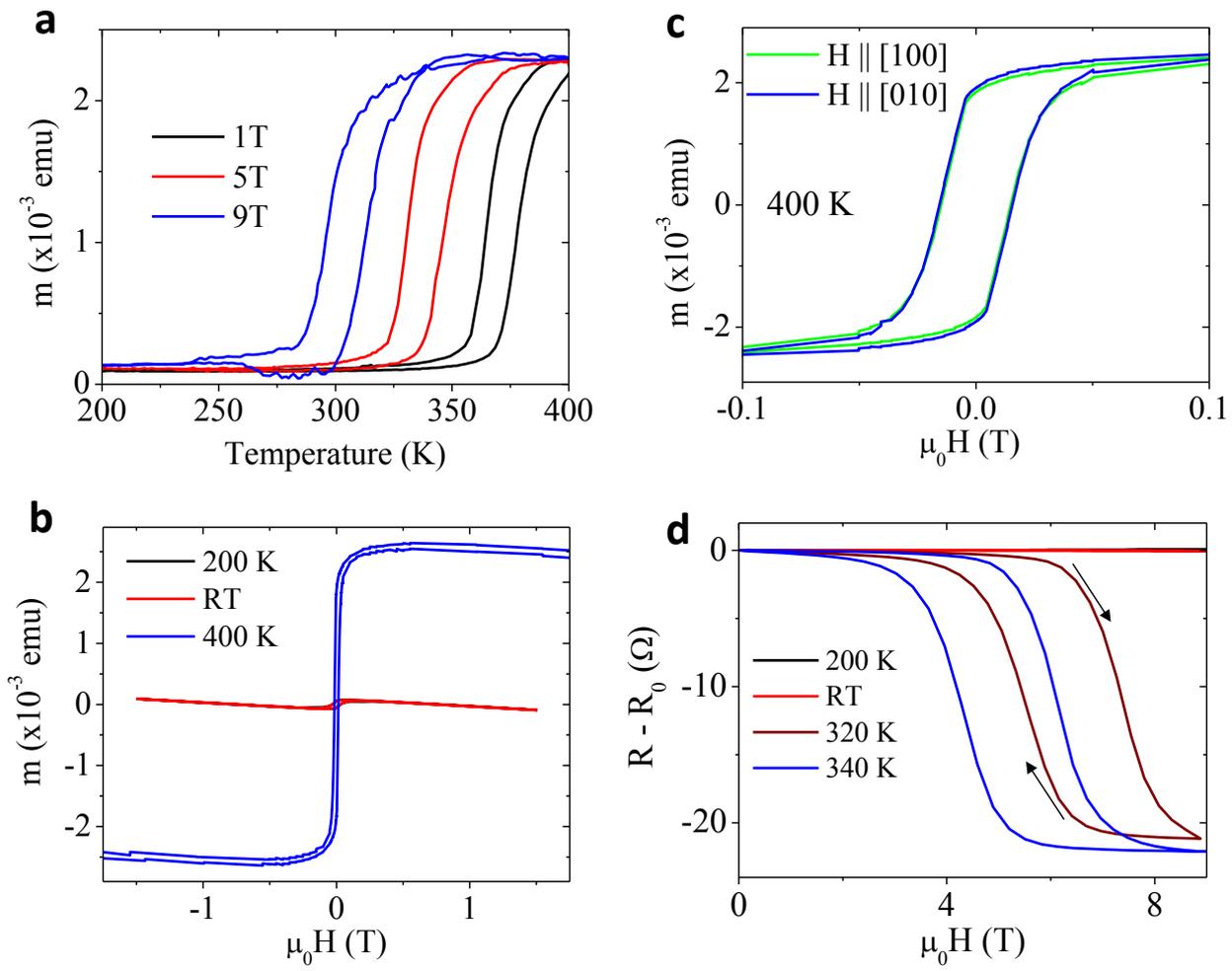

**Figure 3. Magnetic properties. a,** Temperature dependence of the magnetization at 1, 5, and 9 T applied magnetic fields. b, Magnetic field dependence of the magnetization at 200 K, room-temperature, and 400 K. **c,** Comparison of 400 K magnetization loops for magnetic field applied along the [100] and [010] MgO crystallographic directions. d, Magnetic field dependence of the resistance at 200 K, room-temperature, 320 K, and 340 K.

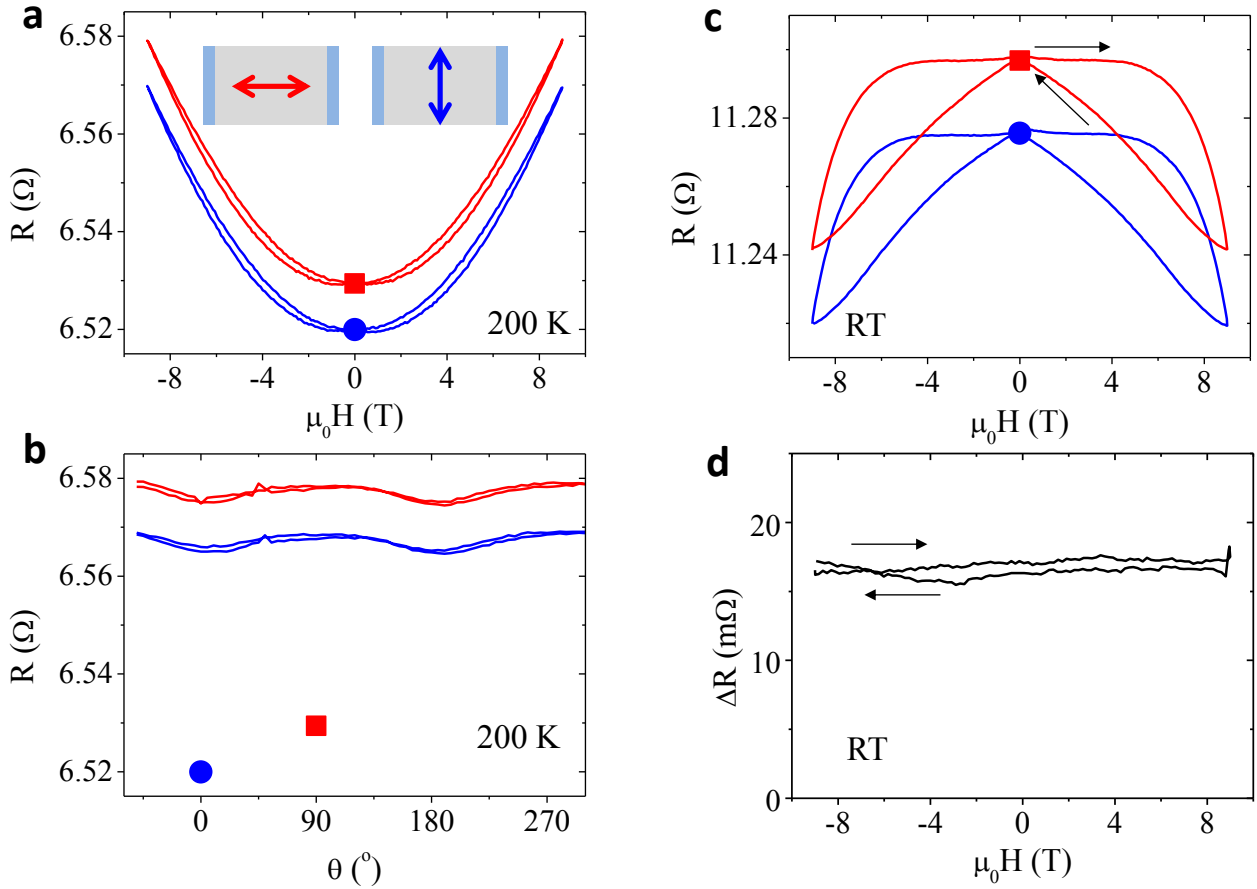

**Figure 4. Stability of the memory states at high magnetic fields. a,** Resistance of the two memory states at 200K and in a magnetic field swept between ±9 T along the [100] direction. **b,** Same as **a,** in a 9 T rotating magnetic field. In **a,** and **b,** the red square and blue circle indicate the zero-field resistances of the two memory states after the cooling-in-field writing procedure. **c,** Same as **a,** measured at room temperature. **d,** The difference between resistances of the two memory states plotted in **c,**.